\documentclass[showpacs,preprintnumbers,showkeys]{revtex4-1}
\usepackage{graphicx}
\usepackage{bm}
\usepackage{amsmath}

\begin{document}

\title{Entanglement Distillation From Gaussian
Input States By Coherent Photon Addition}
\author{Pei Zhang, Hong-Rong Li,\footnote[1]{hrli@mail.xjtu.edu.cn} Hong Gao, and Fu-Li Li}

\address{MOE Key Laboratory for Nonequilibrium Synthesis and
Modulation of Condensed Matter, Department of Applied Physics, Xi'an
Jiaotong University, Xi'an 710049, People's Republic of China}

\begin{abstract}
The entanglement between Gaussian entangled states can be increased by
non-Gaussian operations. We design a new scheme named coherent photon addition, which can coherently
add one photon generated by spontaneous parametric down conversation
process to Gaussian quadrature-entangled light pulses created by a non-degenerate
optical parametric amplifier. This operation can increase the entanglement
of input two-mode Gaussian states as an entanglement distillation,
and provides us a new method of non-Gaussian operation. This scheme
can also help us to study the decoherence
of adding one to two-mode Gaussian states from coherent photon
addition to normal photon addition.

\end{abstract}



\maketitle

% PACS, the Physics and Astronomy
% Classification Scheme.

\section{Introduction}
Quantum states of light are very important resource for quantum
communication and computation protocols \cite{Nilson00,knill01}. Especially,
entangled states play a key role in quantum information processing (QIP),
such as quantum teleportation \cite{ben93}, entanglement swapping \cite%
{zuk93}, remote state preparation \cite{ben01} and quantum repeaters \cite%
{bri98}. Quantum states of propagating light beams can be treated as
discrete single photon states or quantum continuous variables (QCV) states,
which can be analyzed either by photon counting or by homodyne detection,
respectively. QCV is a very interesting system for QIP \cite{bra03}. It
encodes the information in the quadratures $x$ and $p$ of traveling light
fields. In most studies, the QCV states are limited to Gaussian states.
Because the Gaussian states have well understood theoretical structure and can be
easily generated experimentally \cite{bra03}. For example, Gaussian two-mode
squeezed state (QCV entangled light) can be efficiently produced via optical
parametric amplification (OPA), and the two-mode squeezing degree can be
easily varied by adjusting the pump power. However, non-Gaussian operations and
non-Gaussian states are
also very important. It has been shown that Gaussian
entanglement distillation requires non-Gaussian operations \cite{ei02}.
And non-Gaussian states have been employed to fulfil optimal quantum cloning
and minimal disturbance measurement for coherent states \cite{cerf05}.
There have been some proposals to detect entanglement in non-Gaussian
states \cite{agar}. With
the current experiment conditions, one of the practical ways to prepare
non-Gaussian states is the conditional subtraction\ (addition) of photons
from (to) Gaussian entangled beams \cite{sub,add}. And the properties of
these non-Gaussian states have been investigated, including the
non-classicality, inseparability and entanglement \cite{pro,our07,lhr07}.

As we know, entanglement distillation allows one to produce stronger
entanglement between distant sites, which is essential for long-distance
quantum communications \cite{ben96}. In the discrete variables systems,
entanglement distillation has been well studied and demonstrated by several
experiments recently \cite{pan}. While in the QCV systems, several proposals
have been put forward to realize QCV distillation \cite{duan00,ei04}.
Recently, the non-Gaussian operations of photon subtraction, photon addition,
first subtraction then addition (or first addition then subtraction) and coherent photon subtracted (CPS)
have been studied both theoretically and experimentally \cite{pro,our07,lhr07,ping08}.
All these non-Gaussian operations can be used to
generate an entanglement enhanced QCV states as entanglement distillation when
applying it on the two-mode squeezed states.
Different operation has its special property, and all these non-Gaussian operations
are useful to QIP in the QCV systems.
Except the methods mentioned above,
are there some other practicable non-Gaussian operations which can also be used to
increase the entanglement? In this paper, we first describe an experimental scheme
on how to realize a new non-Gaussian operation named coherent photon addition (CPA).
We then give a simple theoretical analysis about this process,
and prove that this operation can also increase entanglement of the initial two-mode Gaussian states.
At the end of this paper, we give a brief discussion and conclusion.

\section{Experimental Scheme}

Inspired by the methods of photon addition and CPS, we design an experimental scheme for CPA process, which can
be readily realized in the lab. The experimental scheme
is shown in Fig. 1. A laser field, e.g. Ti:sapphire laser, is split into two parts by a high-transmission beam splitter
(HT-BS). The small part (reflected part) is used as reference light for detection system, and the main part (transmitted
part) is frequency doubled by a type-I non-critically phase-matched
potassium niobate (KNbO$_{3}$) crystal. One part of the frequency doubled beam pumps a type-II KNbO3 crystal used for OPA interaction process, generating Gaussian quadrature two-mode entangled states spatially separated by an angle of
$10^\circ$ and also with orthogonal polarizations defined
as horizontal (H) polarization and vertical (V) polarization.
Another part of the frequency-doubled beam pumps a type-I
beta-barium borate (BBO) nonlinear crystal generating the correlated
photon pairs. The entangled Gaussian beams are spatially combined on
a polarizing beam splitter (PBS$_1$), but remain different
polarizations. Then this beam is injected in the BBO nonlinear
crystal as a seed field of signal mode, and the conditional
preparation of the photon-added state takes place every time that a
single photon is detected in the correlated idler mode. To achieve
adding one photon to the two mode coherently, we can adjust the
polarization of pump beam of BBO crystal by a half-wave plate (HWP)
in order to generate $45^\circ$-polarized twin photons.
In this case, when the signal beam (mixed with the signal photons and
the entangled Gaussian beams)
pass through the second PBS$_2$, two orthogonal polarized two modes beams will be
reflected or transmitted due to the polarization, and the
$45^\circ$-polarized photons are coherently separated then traveling
with the separated two-mode Gaussian beams.
%The bright local oscillator beams can be also added to the CPA beam on the PBS$_2$ for homodyne detection.
If the detector on the
idler route is clicked, we can get that one photon has been
coherently added to the Gaussian quadrature-entangled beam generated
by OPA.
% The intensities and the relative phase of local oscillator can be easily adjusted by quarter-wave plate (QWP), HWP and piezoelectric transducer (PZT).
At the end of this setup, we can
perform a homodyne detection to analyze the nonclassical character
of the CPA states.

\begin{figure}[tbh]
\includegraphics[width=10cm]{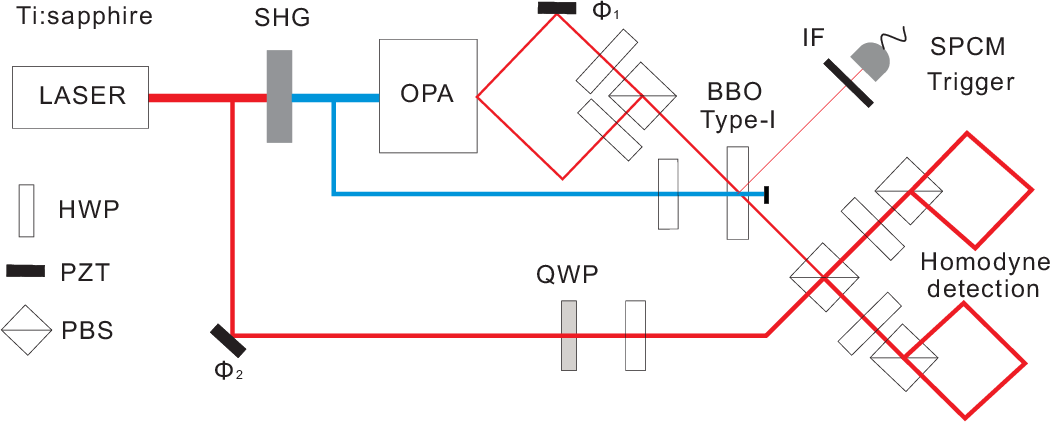}
\caption{Experimental scheme of coherent photon-addition to two-mode
squeezed states. The primary light source is a mode-locked Ti:sapphire laser
whose pulses are split to two spatial modes by a high-transmission beam
splitter (HT-BS), one is used for local oscillator and the other is
frequency-doubled by a second harmonic generation (SHG). The frequency-doubled
beam is used as the pump light
for OPA crystal and type-I BBO crystal. $\Phi_{1}$ and $%
\Phi_{2}$ are varied by the piezoelectric transducer (PZT). All the PBS are
translated vertical polarized light and reflected horizontal polarized
light. F is a combination of spectral and spatial filter made of a
interference filter and a single-spatial-mode optical fiber directly
connected to a single-photon counting module (SPCM). Additional optics and
computer control to realize the homodyne detection and coincidence count are
not shown here for the sake of clarity.}
\end{figure}

The main difference between CPA and previous photon addition \cite{add} is that
CPA process adds one photon to two modes coherently while photon addition adds
one photon determinately to one of two modes.
In this setup, single photons are generated via spontaneous parametric down-conversion (SPDC) process, and combined
with Gaussian quadrature-entangled modes in one spatial routes but with
different polarizations. So we can coherently distribute one photon to separated
entangled state by using a PBS. As we mainly use polarization information of photons and pulses
to control their spatial routes, so the relative phases except $\Phi_{1}$
and $\Phi_{2}$ can be precisely adjusted by wave plates. $\Phi_{1}$ and $%
\Phi_{2}$ are controlled by the PZT, which can precisely
realize a phase scan for experimental request. In this setup, we can
easily adjust the polarization of pump beam of SPDC process to generate
arbitrary polarized entanglement twin-photon states. Especially, when the
twin-photon's polarization is at $45^\circ$, we can
realize the CPA process as mentioned above; when the twin-photon's
polarization is at $0^\circ$ or $90^\circ$, we
can get that one photon is added to one of two modes definitely
(which corresponding to the normal photon addition) \cite{add}. So, this setup
can help us to study the decoherence of adding one photon to
two-mode Gaussian states from CPA to normal photon
addition.

The precision of adding exactly a single photon is mainly
limited by the non-unit detection efficiency of SPCM trigger and the
higher-order emissions from the SPDC process in BBO crystal.
However, these limitations are not very restrictive for the present
application where the down-conversion should be operated in a regime
of low gain.

\section{Theoretical Analysis}

We have presented an experimental scheme for CPA process, and our
proposal can be also used to study the decoherence of photon addition.
We will give a simple analysis to the output states of our scheme and to
see whether the entanglement is increased or not in the following.

The two-mode squeezed state can be expressed in the Fock state basis as
follows
\begin{equation}
\left\vert S(\lambda )\right\rangle =\sqrt{1-\lambda ^2}\displaystyle \sum ^\infty
_{n=0}\lambda ^n \left\vert n,n\right\rangle ,
\end{equation}%
where $\lambda =\tanh (r)$ and $r$ is the squeezing degree
parameter. CPA operation can be written as
\begin{equation}
 U_{CPA} = \sqrt{1-\mu^2}a_1^\dag
 +\mu a_2^\dag.
\end{equation}%
where $\mu $ is a complex parameter, and $0\leq |\mu|\leq 1$. $a_{1}^{\dag }$ and $a_{2}^{\dag }$ are
photon creation operators for modes 1 and 2, respectively. Then we can calculate the CPA states:
\begin{equation}
\left\vert S(\lambda ,1)\right\rangle =U_{CPA}\left\vert S(\lambda )\right\rangle   \nonumber\\
=(1-\lambda ^{2})\displaystyle \sum ^\infty_{n=0}\lambda ^n \sqrt{n+1}(\sqrt{1-\mu^{2}}\left\vert
n+1,n\right\rangle +\mu\left\vert n,n+1\right\rangle ).
\end{equation}%
Especially, when $\mu = \frac{\sqrt{2}}{2}$, the CPA state is
\begin{equation}
\left\vert S(\lambda ,1)\right\rangle_{\mu = \frac{\sqrt{2}}{2}}
=\frac{1-\lambda ^{2}}{\sqrt{2}}\displaystyle \sum ^\infty
_{n=0}\lambda ^{n}\sqrt{n+1}(\left\vert
n+1,n\right\rangle +\left\vert n,n+1\right\rangle ).
\end{equation}%

Compared with the CPS process, we can define the CPS operator as
\begin{equation}
U_{CPS} = \sqrt{1-\mu^{2}}a_{1}+\mu a_{2}.
\end{equation}%
$a_{1}$ and $a_{2}$ are photon annihilation operators for modes 1 and 2, respectively.
So the process of coherently subtracting one photon from a two mode
squeezed state can be written as%
\begin{equation}
\left\vert S(\lambda ,-1)\right\rangle = U_{CPS}\left\vert
S(\lambda )\right\rangle .
\end{equation}%

Without loss generality, we consider the case $\mu = \frac{\sqrt{2}}{2}$. We can get%
\begin{eqnarray}
\left\vert S(\lambda ,-1)\right\rangle_{\mu = \frac{\sqrt{2}}{2}} &=&\frac{1-\lambda ^{2}}{\sqrt{2}}%
\displaystyle \sum ^\infty
_{n=1}\lambda ^{n-1}\sqrt{n}(\left\vert
n-1,n\right\rangle +\left\vert n,n-1\right\rangle )  \nonumber\\
&=&\frac{1-\lambda ^{2}}{\sqrt{2}}\displaystyle \sum ^\infty_{n^{\prime }=0}
\lambda ^{n^{\prime }}\sqrt{n^{\prime }+1}(\left\vert n^{\prime
},n^{\prime }+1\right\rangle +\left\vert n^{\prime }+1,n^{\prime
}\right\rangle ),
\end{eqnarray}%
where $n^{\prime }=n-1$.
The final state of CPS has been experimental tested that the entanglement,
quantified by negativity \cite{vidal02}, is
higher than the initial two-mode squeezed state with up to 3 dB of
squeezing, and even small experimental improvements should
significantly increase this limit \cite{our07}. We can see that Eq. (6) is almost same as
Eq. (3) except for different start point of $n$. So this directly gives us a prediction that
the CPA process can not only generate non-Gaussian states as CPS,
but also can be used as entanglement distillation.

From simple analysis above, we proof in principal that CPA operation can increase the entanglement between Gaussian states. The difference between Eq. (3) and Eq. (6) shows that CPA operation can act on squeezed vacuum state which CPS operation can
not. We know that the squeezed vacuum state is a very useful state, so CPA operation should be more useful than CPA
operation. Furthermore, we may combine CPS and CPA process
together, such as CPS then CPA or CPA then CPS, to study some new
phenomenon.

\section{Conclusion}

In conclusion, we present an experimental scheme of CPA, which allows
us to increase the entanglement between two-mode Gaussian state. This
non-Gaussian operation can generate similar non-Gaussian states with
CPS process, even through the initial state is squeezed vacuum state. And the
realization of CPA makes it possible to do further studies on combining CPS and
CPA. Further more, the coherent parameter $\mu$ of our setup can be easily adjusted by
rotating the polarization of pump pulses of BBO, which make us possible to study the decoherence
property of adding one photon to two-mode Gaussian states from coherent photon
addition to normal photon addition continuously. The final non-Gaussian CPA states can in principle be used as a
starting point for a "Gaussification" procedure \cite{ei04}. The
experimental setup we put forward here is realizable
under current technique. It is useful for Gaussian entanglement
distillation, and it is also one of the key steps for long-distance
quantum communications with QCV.

\section{Acknowledgement }

The authors thank Ai-Ping Fang and Yang Yang for interesting and
helpful discussion. This work is supported by the Fundamental Research Funds for the Central Universities,
the National Fundamental Research Program (2010CB923102) and National Natural Science
Foundation of China (Grant No. 11004158, 10774139, 11074198, and 60778021).

\end{document}